\documentclass[aps,prl,preprint,showpacs,amsmath]{revtex4-1}
\usepackage{graphicx}
\usepackage{bm}
\usepackage{hyperref} 
\begin{document}

\title{Radiation reaction force induced nonlinear mixing of Raman sidebands of an ultra-intense laser pulse in a plasma}

\author{Naveen Kumar}\email{kumar@mpi-hd.mpg.de}
\author{Karen Z. Hatsagortsyan}
\author{Christoph H. Keitel}
\affiliation{Max-Planck-Institut f\"ur Kernphysik, Saupfercheckweg 1, D-69117 Heidelberg, Germany}

\pacs{52.35.-g, 52.40.Mj, 52.65.-y}

\begin{abstract}
Stimulated Raman scattering of an ultra-intense laser pulse in plasmas is studied by perturbatively including the leading order term of the Landau-Lifshitz radiation reaction force in the equation of motion for plasma electrons. In this approximation, radiation reaction force causes phase shift in nonlinear current densities that drive the two Raman sidebands (anti-Stokes and Stokes waves), manifesting itself into the nonlinear mixing of two sidebands. This mixing results in a strong enhancement in the growth of the forward Raman scattering instability. 
\end{abstract}

\maketitle

Parametric instabilities of a laser pulse in a plasma are important due to their applications in the area of laser driven fusion, laser wakefield acceleration in plasmas, and have been investigated for decades \cite{Kruer:2003yq,*Brueckner:1974uq,drake:778,*tripathi:468,*mckinstrie:2626,*Sakharov:1994fk,decker:2047,*Barr:1999ve,*jr.:1440,*guerin:2807,*Quesnel:1997fk,Esarey:2009fk}. Stimulated Raman scattering (SRS) of the laser pulse in plasmas is one of the prominent examples of parametric instabilities carrying significant importance on account of being responsible for the generation of hot electrons in fast ignition fusion \cite{Kruer:2003yq,drake:778}, and strong plasma wakefield excitation in laser driven wakefield acceleration \cite{decker:2047,Esarey:2009fk}. In Raman scattering, the incident laser pump decays either into two forward moving daughter electromagnetic waves (forward Raman scattering) or into a single backward moving daughter electromagnetic wave (backward Raman scattering), and a plasma wave. The daughter waves have their frequencies upshifted (anti-Stokes waves) and downshifted (Stokes wave) from the laser pump by the magnitude which equals the excited plasma wave frequency. Thus the SRS is categorised as a four-wave decay interaction process.

At high laser intensities $I_{l}\sim 10^{19-21}\,\text{W/cm}^2$, the growth rate of the parametric instabilities becomes smaller due to the relativistic Lorentz factor and the rising part of the laser pulse suffers from these instabilities \cite{decker:2047}. However, at ultra-high laser intensities, $I_{l}\ge 10^{22}\,\text{W/cm}^2$, the role of radiation reaction force becomes important too \cite{Di-Piazza:2012uq,Di-Piazza:2009kx,*Schlegel:2012bh,*Sokolov:2010kl,*Chen:2011uq,*Tamburini:2010fk,*Keitel:1998fk}. Such ultra-intense laser systems are on the anvil after the commissioning of the Extreme Light Infrastructure (ELI) project in Europe \cite{eli:kx}. Due to radiation reaction force, the laser suffers damping while propagating in a plasma. This damping of the laser radiation makes the laser pulse vulnerable to the plasma instabilities in following ways: First, as the laser propagates in the plasma, it's effective intensity decreases which lowers the relativistic Lorentz factor and the laser pulse becomes susceptible to parametric instabilities. Second, as the laser looses energy due to radiation reaction force it facilitates, apart from the usual parametric decay processes, the availability of an additional source of free energy for perturbations to grow in the plasma. Third, the phase shift, caused by radiation reaction force, in the nonlinear current densities can mediate the mixing of the scattered daughter electromagnetic waves which can now grow faster utilizing efficiently the additional channel of the laser energy depletion in the plasma caused by radiation reaction force induced damping of the pump laser. This necessitates to include the effect of radiation reaction force in the theoretical formalism of the parametric instabilities in the plasma.

In this Letter, we include the effect of radiation reaction force and study the SRS of an ultra-intense laser in a plasma. We treat the radiation reaction force effects in the classical electrodynamics regime where quantum effects arising due to photon recoil and spin are negligible \cite{Di-Piazza:2012uq}. For this to be valid, the wavelength and magnitude of the external electromagnetic field in the instantaneous rest frame of the electron must satisfy  $\lambda >> \lambda_C,\, E << E_{\text{cr}}$, where $\lambda_C = 3.9 \times 10^{-11}$ cm is the Compton wavelength and $E_{\text{cr}} = 1.3 \times 10^{16}$ V/cm is the critical field of quantum electrodynamics \cite{Di-Piazza:2012uq}. For the laser intensities planned in the ELI project $I_{l}\sim 10^{22-23}\,\text{W/cm}^2$, these two criteria can be fulfilled \cite{eli:kx}. In the classical electrodynamics regime, the Landau-Lifshitz radiation reaction force \cite{Landau:2005fr} correctly describes the equation of motion for a relativistic charged particle \cite {Di-Piazza:2012uq}. We include the leading order term of the Landau-Lifshitz radiation reaction force in the equation of motion for the plasma electrons. We incorporate the radiation reaction force perturbatively to derive the quiver momentum of the electron in the laser field, focusing on the phase slippage caused by radiation reaction force on the quiver momentum of oscillating electrons. On calculating the growth of the SRS, we find that the inclusion of radiation reaction force tends to enhance the growth of the SRS. The growth of the forward Raman scattering (FRS) instability gets strongly enhanced by the nonlinear mixing of the anti-Stokes and the Stokes waves mediated by radiation reaction force in the plasma, which may have important implications for the ELI project \cite{eli:kx}. Though the growth of the backward Raman scattering (BRS) instability, being in the strongly-coupled regime, doesn't experience a strong enhancement due to radiation reaction force. 

We consider the propagation of a circularly polarized (CP) pump laser along the $\hat{z}$ direction in an underdense plasma with uniform plasma electron density $n_0$. The ions are assumed to be at rest. Inclusion of ion motion leads to the appearance of an additional ion mode instability, however it doesn't couple with the SRS instability \cite{max:1480,*Drake:1976fk}. Equation of motion for an electron in the laser electric and magnetic fields including the leading order term of the Landau-Lifshitz radiation reaction force is 
\begin{equation}
\frac{\partial \mathbf{p}}{\partial t} + \bm{\upsilon}\cdot \nabla\mathbf{p}=-e \left(\mathbf{E}+\frac{1}{c}\bm{\upsilon}\times\mathbf{B}\right)-\frac{2e^4}{3m_e^2c^5}\gamma^2\bm{\upsilon}\left[\left(\mathbf{E}+\frac{1}{c}\bm{\upsilon}\times\mathbf{B}\right)^2-\left(\frac{\bm{\upsilon}}{c}\cdot \bm{E}\right)^2\right],
\label{eom}
\end{equation} 
where $\gamma=1/\sqrt{1-\upsilon^2/c^2},\, e$ is the electronic charge, $m_e$ is the electron mass, and $c$ is the velocity of the light in vacuum. The other terms of the Landau-Lifshitz radiation force are $1/\gamma$ times smaller than leading order term \cite {Landau:2005fr} and can be ignored since for the relevant laser intensities required to probe the radiation reaction force induced effects, the Lorentz factor is always large $\gamma \gg 1$.
We first solve this equation of motion by ignoring the radiation reaction term and by expressing the electric and magnetic fields in potentials as $\bm{E}=-\nabla\phi-\partial\bm{A}/ \partial c t,\, \bm{B}=\nabla\times\bm{A}$. In a 1D approximation valid when $r_0\gg\lambda_0$ (where $r_0$ is the spot-size and $\lambda_0$ is the wavelength of the pump laser pulse), this gives \cite{decker:2047,Gibbon:2005ys}
\begin{gather}
\bm{p}_{\perp}=\frac{e}{c}\bm{A},\nonumber \\
\frac{\partial \upsilon_z}{\partial t}=\frac{e\nabla \phi}{m_e\gamma_0} -\frac{e^2}{2m_e^2\gamma_0^2 c^2}\nabla |{A}|^2,
\label{velocityZ}
\end{gather}
where $\gamma_0=(1+a_0^2/2)^{1/2},\, a_0=eA_0/m_ec^2,\bm{A}=\bm{A}_0 e^{i\psi_{0}}/2 + c.c,\,\bm{A}_0=\bm{\sigma}A_{0},\bm{\sigma}=(\hat{\bm x}+i\hat{\bm y})/\sqrt{2},\,\text{and}\,\psi_0=k_0 z-\omega_0 t$. A plane monochromatic CP light has $\nabla |A_0|^2=0$, so it doesn't cause any charge separation leading to no component of velocity in the $\hat{z}$ direction. This is the so-called Akhiezer-Polovin solution for a purely transverse monochromatic CP light in plasmas  \cite{Akhiezer:1956mz,decker:2047,S.V.-Bulanov:2001kx}. However, the scattering of the laser pulse leads to the total vector potential of the form \cite{decker:2047,Gibbon:2005ys}
\begin{gather}
\bm{A}=\frac{1}{2}\left[\bm{A}_0 e^{i\psi_{0}}+\bm{\delta A}_{+}e^{i \bm{k}_{\perp}.\bm{x}_{\perp}} e^{i\psi_{+}} + \bm{\delta A}_{-}^{*}e^{-i\bm{k}_{\perp}.\bm{x}_{\perp}}e^{-i\psi^{*}_{-}}\right]+c.c,
\label{scattering}
\end{gather}
where $\bm{\delta A}_{+}= \bm{\sigma} \delta A_{+}$, $\bm{\delta A}_{-}^{*}=\bm{\sigma} \delta A_{-}^{*}$, $\bm{\delta A}_{+}$ and $\bm{\delta A}_{-}$ represent the anti-Stokes and the Stokes waves respectively, $\psi_{+}=(k_z+k_0)z-(\omega+\omega_0)t,\,\psi_{-}^{*}=(k_z-k_0)z-(\omega^{*}-\omega_0)t$, and $\omega_0,\,k_0$ are the carrier frequency and wavevector of the pump laser respectively. Scattering process in Eq.\eqref{scattering}, for all $\omega$ and $k$, describes modulational interaction in plasmas. For $\omega=\omega_p^{'}\,(\omega_p^{'} < \omega_0)$, where $\omega_p^{'}=\sqrt{\omega_p/\gamma_0},\omega_p = \sqrt {4\pi n_o e^2 /m_e},$ and $k_z \equiv k_p^{'} \approx \omega_p^{'}/c$, the scattering process is known as the SRS and the scattered waves with frequencies (wavevectors) $\omega_p^{'}+\omega_0\, (k_p^{'}+k_0)$, and $\omega_p^{'}-\omega_0\, (k_p^{'}-k_0)$ are the two Raman sidebands, also known as the anti-Stokes and the Stokes waves respectively. Beating of the Stokes and the anti-Stokes waves with the pump laser leads to the density perturbation, $\delta n/ n_0$ (plasma wave oscillations). It can be estimated after solving the equation of continuity and Poisson equation together with Eq.\eqref{velocityZ}, and reads as \cite{decker:2047,Gibbon:2005ys}
\begin{equation}
\delta \tilde{n}=\frac{e^2 k_z^2}{2m_e^2\gamma_0^2c^2 D_e}\left({A}_{0}^{*}{\delta A}_{+}+{A}_{0}{\delta A}_{-}\right),
\label{plasmaosc}
\end{equation}
where $D_e=\omega^2-\omega_p^{'2},\delta n/n_0=\delta \tilde{n}e^{i\psi}e^{i\bm{k}_{\perp}.\bm{x}_{\perp}}/2+c.c,$ and $\psi\equiv \psi_{+}-\psi_0 \equiv \psi_{-}+\psi_0=k_z z -\omega t$, which causes an axial component of velocity and momentum $\beta_z=\upsilon_z/c \ll 1, p_z \ll p_{\perp}$. 

Now we use Eqs.\eqref{velocityZ} and \eqref{scattering} to simplify the radiation reaction term in Eq.\eqref{eom} and solve the full equation of motion  perturbatively to include radiation reaction force.  Writing the CP laser pulse as $\bm{A}=\bm{A_{\perp}}(\bm{x}_{\perp},z,t)e^{i\psi_0}/2+c.c.$, with its amplitude varying slowly i.e. $ \left|\partial \bm{A}_{\perp}/{\partial t}\right|\ll\left|\omega_0 \bm{A}_{\perp}\right|,\, \left|\partial \bm{A}_{\perp}/{\partial z}\right|\ll\left|k_0 \bm{A}_{\perp}\right|$, and $|\phi| \ll |\bm{A}|, \omega_p^2/\gamma \omega_0^2 \ll 1$, and $\gamma=(1+e^2|\bm{A}|^2/m_e^2c^4)^{1/2}$, we get the transverse component of the quiver momentum as
\begin{gather}
\frac{\partial}{\partial t}\left(\bm{p}_{\perp}-\frac{e}{c}\bm{A}\right)=-\frac{e\mu\omega_0}{c}\bm{A}\gamma |\bm{A}|^2(1-2\beta_z),
\label{qmom}
\end{gather}
where $\mu=2e^4\omega_0/3m_e^3c^7$, $\beta_z=(\omega/k_z c)\,\delta \tilde{n}\, e^{i \bm{k}_{\perp}.\bm{x}_{\perp}} e^{i\psi}/2+c.c.$, and we have assumed $|\mu \gamma |\bm{A}|^2| \ll 1$, which is valid for laser intensities $I_{l}\le 10^{23}\,\text{W/cm}^2$, for which the influence of radiation reaction force has to be taken into account. One may also note that we don't consider the effect of radiation reaction on plasma oscillations given by \eqref{plasmaosc}. This is justified since $|\phi| \ll |A|$ and the radiation reaction effects associated with the plasma wave are negligible in the case of the collinear movement of plasma electrons and the plasma wave. One can solve Eq. \eqref{qmom} for the equilibrium and the scattered vector potentials by substituting $\bm{A}$ from Eq.\eqref{scattering} and expressing the transverse component of the quiver momentum in an analogous manner as the vector potential $\bm{A}$  in \eqref{scattering} \emph{e.g.} $\bm{p}_{\perp}=[\bm{p}_0 e^{i\psi_{0}}+\bm{p}_{+}e^{i \bm{k}_{\perp}.\bm{x}_{\perp}} e^{i\psi_{+}} + \bm{p}_{-}^{*}e^{-i\bm{k}_{\perp}.\bm{x}_{\perp}}e^{-i\psi^{*}_{-}}]/2+c.c.$, where $\bm{p}_{+}$ and $\bm{p}_{-}^{*}$ have similar polarizations as the anti-Stokes and the Stokes modes. The wave equation for the vector potential after the density perturbation by the ponderomotive force $n=n_0+\delta n$ becomes
\begin{equation}
\nabla^2\bm{A}-\frac{1}{c^2} \frac{\partial^2 \bm{A}}{\partial t^2}=\frac{\omega_p^{2}}{\gamma c^2}\left( 1+\frac{\delta n}{n_0} \right)\frac{c}{e}\bm{p}_{\perp}.
\label{waveeq}
\end{equation}
One can derive the dispersion relations for the pump and Raman sidebands (anti-Stokes and Stokes waves) by collecting the terms involving $e^{i\psi_0}$ and $e^{\pm i\bm{k}_{\perp} \bm{x}_{\perp}}e^{\pm i\psi_{\pm}}$ on both sides of Eq.\eqref{waveeq}. On collecting the terms containing $e^{i\psi_0}$, Eq.\eqref{waveeq} yields the dispersion relation for the equilibrium vector potential as $\omega_0^2=k_0^2c^2+{\omega_p^{'2}}\left(1-{i\mu}|A_0|^2\gamma_0/2\right)$. Without the radiation reaction term, one recovers the dispersion relation of a CP laser light in plasmas. As it is evident from the dispersion relation, the radiation reaction term causes damping of the pump laser field. This damping can be incorporated either by defining a frequency or a wavenumber shift in the pump laser \footnote{One can also incorporate the radiation reaction term by appropriately modifying the plasma frequency, which essentially implies change in the laser pump wavevector arising due to the it's dispersion in the plasma.}. We define a frequency shift of the form $\omega_0=\omega_{0r}-i\delta\omega_0,\,\delta\omega_0 \ll \omega_{0r}$ with the frequency shift $\delta\omega_0$ being $\delta\omega_0={\omega_{p}^{'2}\varepsilon\gamma_0 a_0^2}/{2\omega_{0r}}$, where $\varepsilon=r_e\omega_{0r}/3c,\,r_e=e^2/m_ec^2$ is the classical radius of the electron and without the loss of generality we have assumed $a_0=a_0^*$. This frequency shift should be less than the growth rate, otherwise the growth of the SRS does not occur and the assumption of the locally constant laser field in deriving the growth rates remains no longer valid.

Similarly  collecting the terms containing $e^{\pm i\psi_{\pm}}e^{\pm i\bm{k}_{\perp}.\bm{x}_{\perp}}$, we get from Eq.\eqref{waveeq}
\begin{gather}
 D_{+}{{\delta A}_{+}}=R_{+}\left({{\delta A}_{+}}+{{\delta A}_{-}}\right),\nonumber \\
 D_{-}{{\delta A}_{-}}=R_{-}\left({{\delta A}_{+}}+{{\delta A}_{-}}\right),
\end{gather} 
where
\begin{gather}
D_{\pm}=(\omega \pm \omega_0)^2-k_{\perp}^2c^2-\omega_{p}^{'2}\left(1- i\varepsilon a_0^2\gamma_0\frac{\omega_0}{\omega\pm\omega_0}\right)-(k_z\pm k_0)^2c^2,\nonumber \\
R_{\pm}=\frac{\omega_p^2 a_0^2}{4\gamma_0^3}\left[\frac{k_z^2 c^2}{D_e}\left(1\mp i\varepsilon a_0^2\gamma_0 + \frac{2 i \varepsilon a_0^2\gamma_0}{k_z c}  \frac{\omega\omega_0}{\omega\pm\omega_0}\right)-\left(1 \mp i \varepsilon a_0^2\gamma_0 \frac{\omega}{\omega\pm \omega_0}+ 4 i \varepsilon \gamma_0^3\frac{\omega_0}{\omega\pm\omega_0}\right)\right].
\label{pertAmpl}
\end{gather}
This yields the dispersion relation
\begin{equation}
 \left(\frac{R_{+}}{D_{+}}+\frac{R_{-}}{D_{-}}\right)=1.
 \label{disp_rel}
\end{equation} 
Due to the presence of the radiation reaction term, coupling between the Stokes and the anti-Stokes modes is modified $(R_+ \neq R_-)$, and this form of dispersion relation differs from the dispersion relation derived before \cite{drake:778,Kruer:2003yq,decker:2047,Gibbon:2005ys}. Without the radiation reaction term $\varepsilon=0,$ and $R_+=R_- \equiv R$, the dispersion relation assumes the same form  as derived before (on taking the non-relativistic limit of the above dispersion relation for the non-relativistic case) \cite{drake:778,Kruer:2003yq,decker:2047,Gibbon:2005ys}.

For calculating the growth rate of the FRS in a low-density plasma, $\omega_{p}^{'} \ll \omega_{0r}$, one has to take into account both the Stokes and the anti-Stokes waves as they both are the resonant modes of the plasma \footnote{we justify retaining both modes in the dispersion relation later by calculating and comparing the frequency mismatch for the anti-Stokes mode with the growth rate of the FRS instability.}. After substituting for the pump laser frequency shift $\delta \omega_0$ and ignoring the finite ${k}_{\perp}$ for the FRS, we get $D_{\pm}= (\omega \pm \omega_{0r})^2-\omega_{p}^{'2}-(k_z\pm k_0)^2c^2 $. On writing $\omega=\omega_p^{'}+i\Gamma_{\text{frs}}$, where $\Gamma_{\text{frs}}$ is the growth rate of the FRS instability,  and assuming that both the sidebands (Stokes and anti-Stokes) are resonant \textit{i.e.} $D_{\pm}\approx (\omega \pm \omega_{0r})^2-\omega_{p}^{'2}-(k_z\pm k_0)^2c^2 = 0$, we have $D_{\pm}\approx 2 i \Gamma_{\text{frs}} (\omega_p^{'}\pm \omega_{0r}),\,D_e\approx 2 i \omega_p^{'} \Gamma_{\text{frs}}$. Substituting these expressions in the dispersion relation and taking $k_z^2 c^2\approx \omega_p^{'2},\, \omega_{p}^{'2} - \omega_{0r}^2\approx -\omega_{0r}^2$, we obtain, in the weakly-coupled regime, $\Gamma_{\text{frs}} \ll \omega_p^{'}$, the growth rate which is well approximated by the following expression
\begin{gather}
\Gamma_{\text{frs}}= -\frac{\omega_{p}^{2}\varepsilon a_0^2}{2\omega_{0r}}\pm \frac{1}{\sqrt{8}}\frac{\omega_{p}^{2} a_0}{\gamma_0^2\omega_{0r}}\text{cos}(\theta/2) \sqrt[4]{\left(1+2 a_0^2 \varepsilon^2\gamma_0^4\right)^2+\varepsilon^2a_0^4\gamma_0^2 \left(\frac{\omega_{0r}}{\omega_{p}^{'}}\right)^2},\nonumber \\
\theta = \text{tan}^{-1}\left(-\frac{\varepsilon a_0^2 \gamma_0 (\omega_{0r}/\omega_{p}^{'})}{(1+2 a_0^2 \varepsilon^2\gamma_0^4)}\right).
\label{frs_growth}
\end{gather} 
\begin{figure}[floatfix]
\includegraphics[width=0.8\textwidth, keepaspectratio]{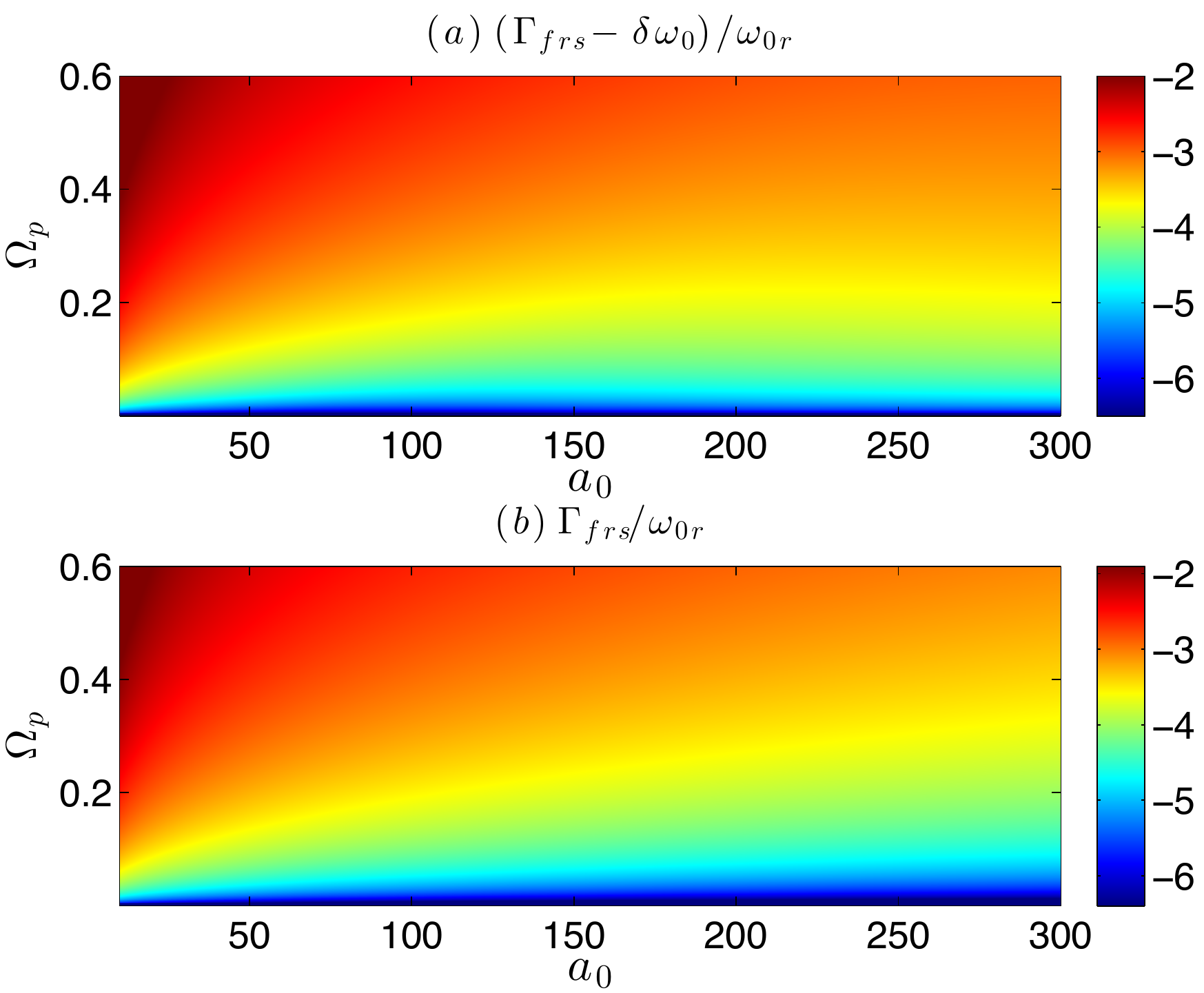}
\caption{(Color online) Normalized growth rate $(\Gamma_{\text{frs}}-\delta\omega_0)/\omega_{0r}$ of the FRS as a function of the normalized plasma density $\Omega_p\equiv \omega_p/\omega_{0r}$ and normalized pump laser amplitude $a_0 = e A_0 / m c^2$ (a) including the radiation reaction force, (b) without the radiation reaction force. Enhancement in the growth rate due to the radiation reaction force is apparent especially at lower plasma densities and higher $a_0$. The normalized growth rate is plotted on Log$_{10}$ scale.}
\label{fig1}
\end{figure}
\noindent
In the case of no radiation reaction force $\varepsilon=0$,  the relativistic growth rate of the FRS instability is same as derived before \cite{decker:2047,Gibbon:2005ys}. Two solutions corresponding to $\pm$ signs represent growing and decaying modes respectively. The decaying mode is damped faster and induces no experimentally detectable signatures in the laser pulse spectrum. The effective growth rate of the FRS instability is $G_{\text{frs}}=\Gamma_{\text{frs}}-\delta \omega_{0}$. It is apparent from the expression that for lower $\omega_p^{'} / \omega_{0r} \ll 1$ and higher laser amplitude $a_0 \gg 1$, radiation reaction force leads to strong enhancement in the growth of the FRS. Fig.\ref{fig1} shows the growth rate of the FRS with (upper panel) and without (lower panel) radiation reaction force. One can immediately notice that radiation reaction force significantly enhances the growth rate of the FRS at higher values of $a_0$. This enhancement is strongest at lower plasma densities and e.g. for $\Omega_p\approx0.02$ and $a_0=300$, there is an order of magnitude enhancement in the growth rate. The enhanced Raman scattering due to radiation reaction force can lead to the laser pulse spectrum broadening centered around the laser carrier frequency. The radiation reaction term also contributes substantially to the growth enhancement of the FRS at higher plasma densities. In this case the growth rate is also higher since it is directly proportional to the square of the plasma frequency. The strong growth enhancement of the FRS instability due to radiation reaction force is counterintuitive as radiation reaction force is generally considered as a damping force similar to collisions in plasmas. This enhancement occurs due to the mixing between the Stokes and anti-Stokes waves as mediated by radiation reaction force. In the absence of radiation reaction force, nonlinear currents that drive the Stokes and the anti-Stokes modes have opposite polarizations and are responsible for the excitation of the respective modes. Due to opposite polarizations of the Stokes and the anti-Stokes modes, the phase shift induced by radiation reaction force - as seen from Eq.\eqref{pertAmpl} -  is opposite for these modes, and rather than cancelling itself gets accumulated in Eq.\eqref{disp_rel}. It is a consequence of the rotation of the electric field vectors of the two modes due to the radiation reaction force induced phase shift, facilitating interaction between the nonlinear current terms in Eq.\eqref{disp_rel}. We term the nonlinear mixing of the modes due to radiation reaction force as the manifestation of this accumulation of the phase shifts. It leads to the enhanced growth rate of the FRS instability in the plasma. One can also intuitively imagine this growth enhancement occurring due to the availability of an additional channel of laser energy decay caused by radiation reaction force which is used up by the FRS instability efficiently when both the Stokes and the anti-Stokes modes are the resonant modes of the plasma. This is usually the case for the FRS as the plasma wave wavevector is much smaller than the pump laser wavevector, i.e. $k_z \ll k_0$ in a low-density plasma. Since, this growth enhancement depends strongly on the excitation of both the Stokes and the anti-Stokes modes in the plasma, it is instructive to estimate the conditions under which both the Stokes and the anti-Stokes modes are excited and also to see if radiation reaction term enhances the growth of the FRS even when only the Stokes mode is excited in the plasma. Kinematical considerations always allow excitation of the Stokes mode $(D_{-}=0)$ in the plasma, however only in a tenuous plasma $(\omega_p^{'}\ll \omega_{0r})$, both the Stokes and the anti-Stokes modes can simultaneously be excited. Assuming that the Stokes mode is excited, one can calculate the frequency mismatch for the anti-Stokes mode which is defined as $\Delta \omega_m = \omega_p^{'}+\omega_{0r} -\left[\omega_p^{'2}+ c^2 (k+k_0)^2 + D_{+}\right]^{1/2}$ and it turns out to be $\Delta\omega_m=-\omega_p^{'3}/\omega_{0r}^2+9 \omega_p^{'4}/4 \omega_{0r}^3$. If this frequency mismatch is smaller than the growth rate $\Gamma_{\text{frs}}-\delta\omega_0$ of the FRS instability, then one has to retain both modes in the dispersion relation while deriving the growth rate of the FRS. Fig.\ref{fig2} depicts the frequency mismatch normalized by the growth rate of the FRS, $|\Delta \omega_m/(\Gamma_{\text{frs}}-\delta\omega_0)|$, with $a_0$ and $\Omega_p \equiv \omega_p/\omega_{0r}$. One can clearly see that the frequency mismatch for the anti-Stokes mode is usually smaller than the growth rate of the FRS for all values of $\Omega_p$ and $a_0$. This necessitates including both the Stokes and the anti-Stokes modes in the analysis of the FRS. The frequency mismatch is indeed much smaller than the growth rate at lower plasma densities and higher $a_0$. This is also the parameter regime where strong enhancement to the FRS growth rate occurs. Hence both the Stokes and the anti-Stokes modes are excited in the plasma leading to strong enhancement of the FRS instability due to radiation reaction force. If one considers only the Stokes mode in the dispersion relation, the growth rate enhancement due to radiation reaction force is marginal as the nonlinear  mixing of the Stokes and the anti-Stokes modes is absent in this case. The growth rate enhancement in this case occurs due to the phase shift caused by radiation reaction force which maintains the laser energy transfer to the Stokes mode for a longer time.
\begin{figure}
\includegraphics[width=0.7\textwidth,keepaspectratio]{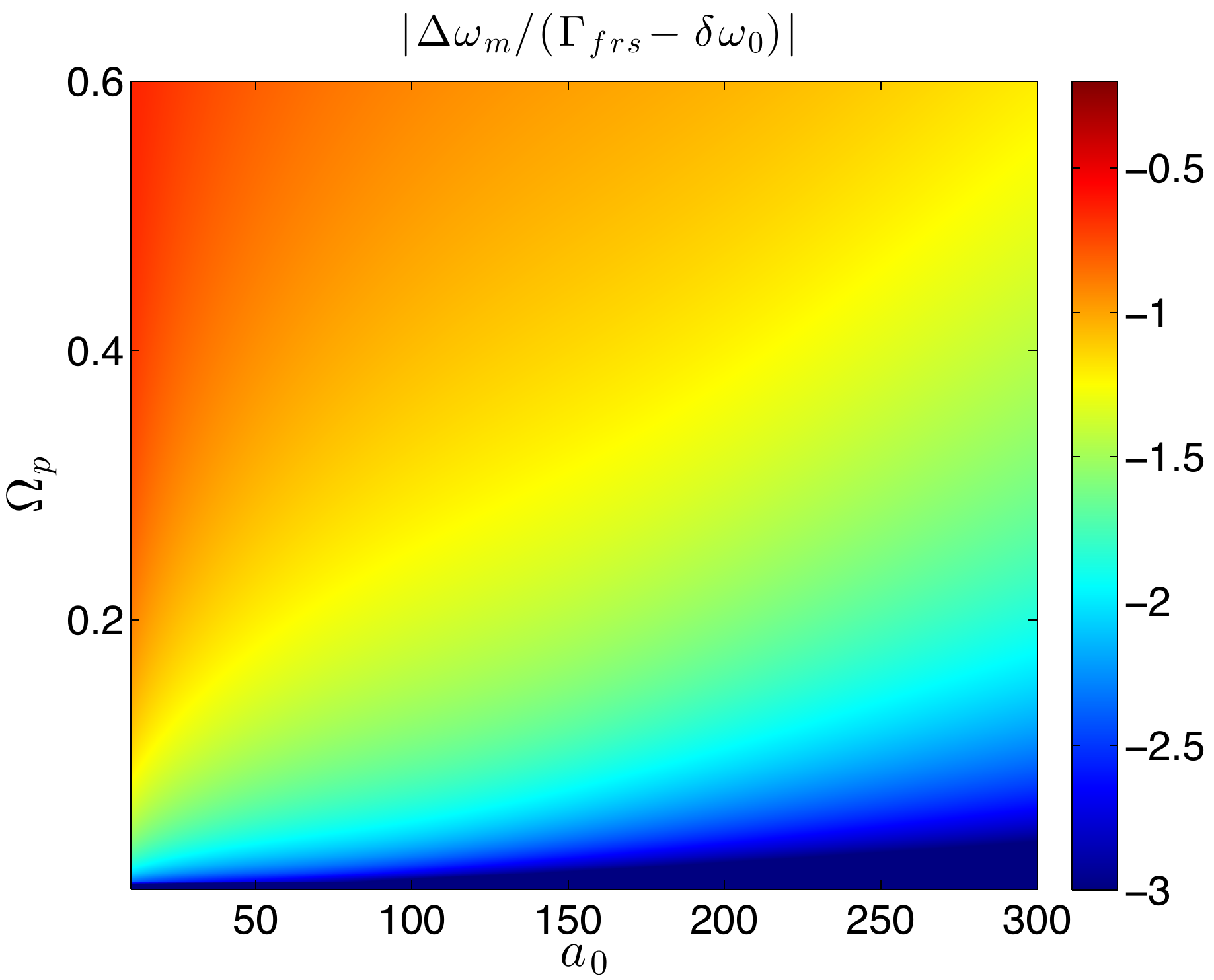}
\caption{(Color online) Normalized frequency mismatch of the anti-Stokes wave ($|\Delta \omega_m /(\Gamma_{\text{frs}}-\delta\omega_0)|$) as a function of normalized plasma frequency $\Omega_p\equiv \omega_p/\omega_{0r}$ and normalized laser amplitude $a_0 = e A_0/m c^2$. One can clearly see that the growth rate $\Gamma_{\text{frs}}-\delta\omega_0$ is larger than the frequency mismatch $|\Delta \omega_m|$, necessitating the need to consider both the Stokes and the anti-Stokes modes in the dispersion relation \eqref{disp_rel} while calculating the growth of the forward Raman scattering. The normalized frequency mismatch is plotted on Log$_{10}$ scale.}
\label{fig2}
\end{figure}
The BRS is essentially a three-wave decay process as the anti-Stokes wave is not the resonant mode of the plasma. For the BRS, we have $k_z \simeq 2 k_0$ and the instability is always in the strongly-coupled regime \textit {i.e.} $\Gamma_{\text{brs}} \gg \omega_{p}^{'}$ (but $\Gamma_{\text{brs}} \ll \omega_{0r}$). One can expand $D_e$ and $D_{-}$ as $D_e\approx -\Gamma_{\text{brs}}^2,\,D_{-}\approx - 2 i \Gamma_{\text{brs}} \omega_{0r}$, and we get the growth rate of the BRS  
\begin{equation}
\Gamma_{\text{brs}}=\frac{\sqrt{3}}{2}\left(\frac{\omega_{0r}}{2\omega_p}\right)^{1/3}\frac{\omega_p a_0^{2/3}}{(1+a_0^2/2)^{1/2}}\left(1+\frac{\varepsilon a_0^2 \gamma_0}{3\sqrt{3}}\right).
\end{equation}
The effective growth rate of the BRS instability is, $G_{\text{brs}}=\Gamma_{\text{brs}}-\delta \omega_{0}$. The radiation reaction term enhances the growth rate of the BRS, however the enhancement due to radiation reaction force is a minor one due to the growth being in the strongly-coupled regime. Unlike the case of the FRS, radiation reaction force does not enhance the growth of the BRS instability strongly as no mixing is possible between the anti-Stokes and the Stokes modes in this case due to the absence of the resonant excitation of the anti-Stokes mode in the plasma.  Since the daughter electromagnetic wave moves in the opposite direction, it also can't exploit the additional pump energy depletion channel caused by the radiation damping of the pump laser field for a longer time. Again for $\varepsilon =0$, one recovers the known growth rate of the BRS \cite{decker:2047,Gibbon:2005ys}. 

We have investigated the influence of the leading order term of the Landau-Lifshitz radiation reaction force on the growth of parametric instabilities namely the SRS in plasmas. It is found that the inclusion of radiation reaction force strongly enhances the growth of the FRS only when both the Stokes and the anti-Stokes modes are the resonant modes of the plasma. The growths of the FRS - with only the Stokes wave excitation - and the BRS are also enhanced by the inclusion of the radiation reaction force though the enhancement is a minor one due to the absence of the radiation reaction force induced nonlinear mixing of the modes. In general, radiation reaction force appears to strongly enhance the growth rate of the SRS involving four-wave decay interaction. These results are important for the ELI project as the ultra-intense laser pulses are expected to create a dense plasma by strongly ionizing the ambient air and also by producing the electron-positron pairs. The subsequent interaction of this plasma with the laser pulse can lead to the onset of parametric instabilities again, now counterintuitively due to radiation reaction force, leading to significant change in the frequency spectra and the shapes of these extremely intense short laser pulses.


\begin{thebibliography}{28}%
\makeatletter
\providecommand \@ifxundefined [1]{%
 \@ifx{#1\undefined}
}%
\providecommand \@ifnum [1]{%
 \ifnum #1\expandafter \@firstoftwo
 \else \expandafter \@secondoftwo
 \fi
}%
\providecommand \@ifx [1]{%
 \ifx #1\expandafter \@firstoftwo
 \else \expandafter \@secondoftwo
 \fi
}%
\providecommand \natexlab [1]{#1}%
\providecommand \enquote  [1]{``#1''}%
\providecommand \bibnamefont  [1]{#1}%
\providecommand \bibfnamefont [1]{#1}%
\providecommand \citenamefont [1]{#1}%
\providecommand \href@noop [0]{\@secondoftwo}%
\providecommand \href [0]{\begingroup \@sanitize@url \@href}%
\providecommand \@href[1]{\@@startlink{#1}\@@href}%
\providecommand \@@href[1]{\endgroup#1\@@endlink}%
\providecommand \@sanitize@url [0]{\catcode `\\12\catcode `\$12\catcode
  `\&12\catcode `\#12\catcode `\^12\catcode `\_12\catcode `\%12\relax}%
\providecommand \@@startlink[1]{}%
\providecommand \@@endlink[0]{}%
\providecommand \url  [0]{\begingroup\@sanitize@url \@url }%
\providecommand \@url [1]{\endgroup\@href {#1}{\urlprefix }}%
\providecommand \urlprefix  [0]{URL }%
\providecommand \Eprint [0]{\href }%
\providecommand \doibase [0]{http://dx.doi.org/}%
\providecommand \selectlanguage [0]{\@gobble}%
\providecommand \bibinfo  [0]{\@secondoftwo}%
\providecommand \bibfield  [0]{\@secondoftwo}%
\providecommand \translation [1]{[#1]}%
\providecommand \BibitemOpen [0]{}%
\providecommand \bibitemStop [0]{}%
\providecommand \bibitemNoStop [0]{.\EOS\space}%
\providecommand \EOS [0]{\spacefactor3000\relax}%
\providecommand \BibitemShut  [1]{\csname bibitem#1\endcsname}%
\let\auto@bib@innerbib\@empty
\bibitem [{\citenamefont {Kruer}(2003)}]{Kruer:2003yq}%
  \BibitemOpen
  \bibfield  {author} {\bibinfo {author} {\bibfnamefont {W.}~\bibnamefont
  {Kruer}},\ }\href@noop {} {\emph {\bibinfo {title} {The Physics Of Laser
  Plasma Interactions (Frontiers in Physics)}}}\ (\bibinfo  {publisher}
  {Westview Press},\ \bibinfo {year} {2003})\BibitemShut {NoStop}%
\bibitem [{\citenamefont {Brueckner}\ and\ \citenamefont
  {Jorna}(1974)}]{Brueckner:1974uq}%
  \BibitemOpen
  \bibfield  {author} {\bibinfo {author} {\bibfnamefont {K.~A.}\ \bibnamefont
  {Brueckner}}\ and\ \bibinfo {author} {\bibfnamefont {S.}~\bibnamefont
  {Jorna}},\ }\href {\doibase 10.1103/RevModPhys.46.325} {\bibfield  {journal}
  {\bibinfo  {journal} {Rev. Mod. Phys.}\ }\textbf {\bibinfo {volume} {46}},\
  \bibinfo {pages} {325} (\bibinfo {year} {1974})}\BibitemShut {NoStop}%
\bibitem [{\citenamefont {Drake}\ \emph {et~al.}(1974)\citenamefont {Drake},
  \citenamefont {Kaw}, \citenamefont {Lee}, \citenamefont {Schmid},
  \citenamefont {Liu},\ and\ \citenamefont {Rosenbluth}}]{drake:778}%
  \BibitemOpen
  \bibfield  {author} {\bibinfo {author} {\bibfnamefont {J.~F.}\ \bibnamefont
  {Drake}}, \bibinfo {author} {\bibfnamefont {P.~K.}\ \bibnamefont {Kaw}},
  \bibinfo {author} {\bibfnamefont {Y.~C.}\ \bibnamefont {Lee}}, \bibinfo
  {author} {\bibfnamefont {G.}~\bibnamefont {Schmid}}, \bibinfo {author}
  {\bibfnamefont {C.~S.}\ \bibnamefont {Liu}}, \ and\ \bibinfo {author}
  {\bibfnamefont {M.~N.}\ \bibnamefont {Rosenbluth}},\ }\href {\doibase
  10.1063/1.1694789} {\bibfield  {journal} {\bibinfo  {journal} {Physics of
  Fluids}\ }\textbf {\bibinfo {volume} {17}},\ \bibinfo {pages} {778} (\bibinfo
  {year} {1974})}\BibitemShut {NoStop}%
\bibitem [{\citenamefont {Tripathi}\ and\ \citenamefont
  {Liu}(1991)}]{tripathi:468}%
  \BibitemOpen
  \bibfield  {author} {\bibinfo {author} {\bibfnamefont {V.~K.}\ \bibnamefont
  {Tripathi}}\ and\ \bibinfo {author} {\bibfnamefont {C.~S.}\ \bibnamefont
  {Liu}},\ }\href {\doibase 10.1063/1.859889} {\bibfield  {journal} {\bibinfo
  {journal} {Physics of Fluids B: Plasma Physics}\ }\textbf {\bibinfo {volume}
  {3}},\ \bibinfo {pages} {468} (\bibinfo {year} {1991})}\BibitemShut {NoStop}%
\bibitem [{\citenamefont {McKinstrie}\ and\ \citenamefont
  {Bingham}(1992)}]{mckinstrie:2626}%
  \BibitemOpen
  \bibfield  {author} {\bibinfo {author} {\bibfnamefont {C.~J.}\ \bibnamefont
  {McKinstrie}}\ and\ \bibinfo {author} {\bibfnamefont {R.}~\bibnamefont
  {Bingham}},\ }\href {\doibase 10.1063/1.860178} {\bibfield  {journal}
  {\bibinfo  {journal} {Physics of Fluids B: Plasma Physics}\ }\textbf
  {\bibinfo {volume} {4}},\ \bibinfo {pages} {2626} (\bibinfo {year}
  {1992})}\BibitemShut {NoStop}%
\bibitem [{\citenamefont {Sakharov}\ and\ \citenamefont
  {Kirsanov}(1994)}]{Sakharov:1994fk}%
  \BibitemOpen
  \bibfield  {author} {\bibinfo {author} {\bibfnamefont {A.~S.}\ \bibnamefont
  {Sakharov}}\ and\ \bibinfo {author} {\bibfnamefont {V.~I.}\ \bibnamefont
  {Kirsanov}},\ }\href {\doibase 10.1103/PhysRevE.49.3274} {\bibfield
  {journal} {\bibinfo  {journal} {Phys. Rev. E}\ }\textbf {\bibinfo {volume}
  {49}},\ \bibinfo {pages} {3274} (\bibinfo {year} {1994})}\BibitemShut
  {NoStop}%
\bibitem [{\citenamefont {Decker}\ \emph {et~al.}(1996)\citenamefont {Decker},
  \citenamefont {Mori}, \citenamefont {Tzeng},\ and\ \citenamefont
  {Katsouleas}}]{decker:2047}%
  \BibitemOpen
  \bibfield  {author} {\bibinfo {author} {\bibfnamefont {C.~D.}\ \bibnamefont
  {Decker}}, \bibinfo {author} {\bibfnamefont {W.~B.}\ \bibnamefont {Mori}},
  \bibinfo {author} {\bibfnamefont {K.-C.}\ \bibnamefont {Tzeng}}, \ and\
  \bibinfo {author} {\bibfnamefont {T.}~\bibnamefont {Katsouleas}},\ }\href
  {\doibase 10.1063/1.872001} {\bibfield  {journal} {\bibinfo  {journal}
  {Physics of Plasmas}\ }\textbf {\bibinfo {volume} {3}},\ \bibinfo {pages}
  {2047} (\bibinfo {year} {1996})}\BibitemShut {NoStop}%
\bibitem [{\citenamefont {Barr}\ \emph {et~al.}(1999)\citenamefont {Barr},
  \citenamefont {Mason},\ and\ \citenamefont {Parr}}]{Barr:1999ve}%
  \BibitemOpen
  \bibfield  {author} {\bibinfo {author} {\bibfnamefont {H.~C.}\ \bibnamefont
  {Barr}}, \bibinfo {author} {\bibfnamefont {P.}~\bibnamefont {Mason}}, \ and\
  \bibinfo {author} {\bibfnamefont {D.~M.}\ \bibnamefont {Parr}},\ }\href
  {\doibase 10.1103/PhysRevLett.83.1606} {\bibfield  {journal} {\bibinfo
  {journal} {Phys. Rev. Lett.}\ }\textbf {\bibinfo {volume} {83}},\ \bibinfo
  {pages} {1606} (\bibinfo {year} {1999})}\BibitemShut {NoStop}%
\bibitem [{\citenamefont {T.~M.~Antonsen}\ and\ \citenamefont
  {Mora}(1993)}]{jr.:1440}%
  \BibitemOpen
  \bibfield  {author} {\bibinfo {author} {\bibfnamefont {J.}~\bibnamefont
  {T.~M.~Antonsen}}\ and\ \bibinfo {author} {\bibfnamefont {P.}~\bibnamefont
  {Mora}},\ }\href {\doibase 10.1063/1.860884} {\bibfield  {journal} {\bibinfo
  {journal} {Physics of Fluids B: Plasma Physics}\ }\textbf {\bibinfo {volume}
  {5}},\ \bibinfo {pages} {1440} (\bibinfo {year} {1993})}\BibitemShut
  {NoStop}%
\bibitem [{\citenamefont {Guerin}\ \emph {et~al.}(1995)\citenamefont {Guerin},
  \citenamefont {Laval}, \citenamefont {Mora}, \citenamefont {Adam},
  \citenamefont {Heron},\ and\ \citenamefont {Bendib}}]{guerin:2807}%
  \BibitemOpen
  \bibfield  {author} {\bibinfo {author} {\bibfnamefont {S.}~\bibnamefont
  {Guerin}}, \bibinfo {author} {\bibfnamefont {G.}~\bibnamefont {Laval}},
  \bibinfo {author} {\bibfnamefont {P.}~\bibnamefont {Mora}}, \bibinfo {author}
  {\bibfnamefont {J.~C.}\ \bibnamefont {Adam}}, \bibinfo {author}
  {\bibfnamefont {A.}~\bibnamefont {Heron}}, \ and\ \bibinfo {author}
  {\bibfnamefont {A.}~\bibnamefont {Bendib}},\ }\href {\doibase
  10.1063/1.871178} {\bibfield  {journal} {\bibinfo  {journal} {Physics of
  Plasmas}\ }\textbf {\bibinfo {volume} {2}},\ \bibinfo {pages} {2807}
  (\bibinfo {year} {1995})}\BibitemShut {NoStop}%
\bibitem [{\citenamefont {Quesnel}\ \emph {et~al.}(1997)\citenamefont
  {Quesnel}, \citenamefont {Mora}, \citenamefont {Adam}, \citenamefont
  {Gu\'erin}, \citenamefont {H\'eron},\ and\ \citenamefont
  {Laval}}]{Quesnel:1997fk}%
  \BibitemOpen
  \bibfield  {author} {\bibinfo {author} {\bibfnamefont {B.}~\bibnamefont
  {Quesnel}}, \bibinfo {author} {\bibfnamefont {P.}~\bibnamefont {Mora}},
  \bibinfo {author} {\bibfnamefont {J.~C.}\ \bibnamefont {Adam}}, \bibinfo
  {author} {\bibfnamefont {S.}~\bibnamefont {Gu\'erin}}, \bibinfo {author}
  {\bibfnamefont {A.}~\bibnamefont {H\'eron}}, \ and\ \bibinfo {author}
  {\bibfnamefont {G.}~\bibnamefont {Laval}},\ }\href {\doibase
  10.1103/PhysRevLett.78.2132} {\bibfield  {journal} {\bibinfo  {journal}
  {Phys. Rev. Lett.}\ }\textbf {\bibinfo {volume} {78}},\ \bibinfo {pages}
  {2132} (\bibinfo {year} {1997})}\BibitemShut {NoStop}%
\bibitem [{\citenamefont {Esarey}\ \emph {et~al.}(2009)\citenamefont {Esarey},
  \citenamefont {Schroeder},\ and\ \citenamefont {Leemans}}]{Esarey:2009fk}%
  \BibitemOpen
  \bibfield  {author} {\bibinfo {author} {\bibfnamefont {E.}~\bibnamefont
  {Esarey}}, \bibinfo {author} {\bibfnamefont {C.~B.}\ \bibnamefont
  {Schroeder}}, \ and\ \bibinfo {author} {\bibfnamefont {W.~P.}\ \bibnamefont
  {Leemans}},\ }\href {\doibase 10.1103/RevModPhys.81.1229} {\bibfield
  {journal} {\bibinfo  {journal} {Rev. Mod. Phys.}\ }\textbf {\bibinfo {volume}
  {81}},\ \bibinfo {pages} {1229} (\bibinfo {year} {2009})}\BibitemShut
  {NoStop}%
\bibitem [{\citenamefont {Di~Piazza}\ \emph {et~al.}(2012)\citenamefont
  {Di~Piazza}, \citenamefont {M\"uller}, \citenamefont {Hatsagortsyan},\ and\
  \citenamefont {Keitel}}]{Di-Piazza:2012uq}%
  \BibitemOpen
  \bibfield  {author} {\bibinfo {author} {\bibfnamefont {A.}~\bibnamefont
  {Di~Piazza}}, \bibinfo {author} {\bibfnamefont {C.}~\bibnamefont {M\"uller}},
  \bibinfo {author} {\bibfnamefont {K.~Z.}\ \bibnamefont {Hatsagortsyan}}, \
  and\ \bibinfo {author} {\bibfnamefont {C.~H.}\ \bibnamefont {Keitel}},\
  }\href {\doibase 10.1103/RevModPhys.84.1177} {\bibfield  {journal} {\bibinfo
  {journal} {Rev. Mod. Phys.}\ }\textbf {\bibinfo {volume} {84}},\ \bibinfo
  {pages} {1177} (\bibinfo {year} {2012})}\BibitemShut {NoStop}%
\bibitem [{\citenamefont {Di~Piazza}\ \emph {et~al.}(2009)\citenamefont
  {Di~Piazza}, \citenamefont {Hatsagortsyan},\ and\ \citenamefont
  {Keitel}}]{Di-Piazza:2009kx}%
  \BibitemOpen
  \bibfield  {author} {\bibinfo {author} {\bibfnamefont {A.}~\bibnamefont
  {Di~Piazza}}, \bibinfo {author} {\bibfnamefont {K.~Z.}\ \bibnamefont
  {Hatsagortsyan}}, \ and\ \bibinfo {author} {\bibfnamefont {C.~H.}\
  \bibnamefont {Keitel}},\ }\href {\doibase 10.1103/PhysRevLett.102.254802}
  {\bibfield  {journal} {\bibinfo  {journal} {Phys. Rev. Lett.}\ }\textbf
  {\bibinfo {volume} {102}},\ \bibinfo {pages} {254802} (\bibinfo {year}
  {2009})}\BibitemShut {NoStop}%
\bibitem [{\citenamefont {Schlegel}\ and\ \citenamefont
  {Tikhonchuk}(2012)}]{Schlegel:2012bh}%
  \BibitemOpen
  \bibfield  {author} {\bibinfo {author} {\bibfnamefont {T.}~\bibnamefont
  {Schlegel}}\ and\ \bibinfo {author} {\bibfnamefont {V.~T.}\ \bibnamefont
  {Tikhonchuk}},\ }\href {http://stacks.iop.org/1367-2630/14/i=7/a=073034}
  {\bibfield  {journal} {\bibinfo  {journal} {New Journal of Physics}\ }\textbf
  {\bibinfo {volume} {14}},\ \bibinfo {pages} {073034} (\bibinfo {year}
  {2012})}\BibitemShut {NoStop}%
\bibitem [{\citenamefont {Sokolov}\ \emph {et~al.}(2010)\citenamefont
  {Sokolov}, \citenamefont {Nees}, \citenamefont {Yanovsky}, \citenamefont
  {Naumova},\ and\ \citenamefont {Mourou}}]{Sokolov:2010kl}%
  \BibitemOpen
  \bibfield  {author} {\bibinfo {author} {\bibfnamefont {I.~V.}\ \bibnamefont
  {Sokolov}}, \bibinfo {author} {\bibfnamefont {J.~A.}\ \bibnamefont {Nees}},
  \bibinfo {author} {\bibfnamefont {V.~P.}\ \bibnamefont {Yanovsky}}, \bibinfo
  {author} {\bibfnamefont {N.~M.}\ \bibnamefont {Naumova}}, \ and\ \bibinfo
  {author} {\bibfnamefont {G.~A.}\ \bibnamefont {Mourou}},\ }\href {\doibase
  10.1103/PhysRevE.81.036412} {\bibfield  {journal} {\bibinfo  {journal} {Phys.
  Rev. E}\ }\textbf {\bibinfo {volume} {81}},\ \bibinfo {pages} {036412}
  (\bibinfo {year} {2010})}\BibitemShut {NoStop}%
\bibitem [{\citenamefont {Chen}\ \emph {et~al.}(2011)\citenamefont {Chen},
  \citenamefont {Pukhov}, \citenamefont {Yu},\ and\ \citenamefont
  {Sheng}}]{Chen:2011uq}%
  \BibitemOpen
  \bibfield  {author} {\bibinfo {author} {\bibfnamefont {M.}~\bibnamefont
  {Chen}}, \bibinfo {author} {\bibfnamefont {A.}~\bibnamefont {Pukhov}},
  \bibinfo {author} {\bibfnamefont {T.-P.}\ \bibnamefont {Yu}}, \ and\ \bibinfo
  {author} {\bibfnamefont {Z.-M.}\ \bibnamefont {Sheng}},\ }\href
  {http://stacks.iop.org/0741-3335/53/i=1/a=014004} {\bibfield  {journal}
  {\bibinfo  {journal} {Plasma Physics and Controlled Fusion}\ }\textbf
  {\bibinfo {volume} {53}},\ \bibinfo {pages} {014004} (\bibinfo {year}
  {2011})}\BibitemShut {NoStop}%
\bibitem [{\citenamefont {Tamburini}\ \emph {et~al.}(2010)\citenamefont
  {Tamburini}, \citenamefont {Pegoraro}, \citenamefont {Piazza}, \citenamefont
  {Keitel},\ and\ \citenamefont {Macchi}}]{Tamburini:2010fk}%
  \BibitemOpen
  \bibfield  {author} {\bibinfo {author} {\bibfnamefont {M.}~\bibnamefont
  {Tamburini}}, \bibinfo {author} {\bibfnamefont {F.}~\bibnamefont {Pegoraro}},
  \bibinfo {author} {\bibfnamefont {A.~Di.}\ \bibnamefont {Piazza}}, \bibinfo
  {author} {\bibfnamefont {C.~H.}\ \bibnamefont {Keitel}}, \ and\ \bibinfo
  {author} {\bibfnamefont {A.}~\bibnamefont {Macchi}},\ }\href
  {http://stacks.iop.org/1367-2630/12/i=12/a=123005} {\bibfield  {journal}
  {\bibinfo  {journal} {New Journal of Physics}\ }\textbf {\bibinfo {volume}
  {12}},\ \bibinfo {pages} {123005} (\bibinfo {year} {2010})}\BibitemShut
  {NoStop}%
\bibitem [{\citenamefont {Keitel}\ \emph {et~al.}(1998)\citenamefont {Keitel},
  \citenamefont {Szymanowski}, \citenamefont {Knight},\ and\ \citenamefont
  {Maquet}}]{Keitel:1998fk}%
  \BibitemOpen
  \bibfield  {author} {\bibinfo {author} {\bibfnamefont {C.~H.}\ \bibnamefont
  {Keitel}}, \bibinfo {author} {\bibfnamefont {C.}~\bibnamefont {Szymanowski}},
  \bibinfo {author} {\bibfnamefont {P.~L.}\ \bibnamefont {Knight}}, \ and\
  \bibinfo {author} {\bibfnamefont {A.}~\bibnamefont {Maquet}},\ }\href
  {http://stacks.iop.org/0953-4075/31/i=3/a=002} {\bibfield  {journal}
  {\bibinfo  {journal} {Journal of Physics B: Atomic, Molecular and Optical
  Physics}\ }\textbf {\bibinfo {volume} {31}},\ \bibinfo {pages} {L75}
  (\bibinfo {year} {1998})}\BibitemShut {NoStop}%
\bibitem [{eli()}]{eli:kx}%
  \BibitemOpen
  \href {http://www.extreme-light-infrastructure.eu} {\enquote {\bibinfo
  {title} {The extreme light infrastructure project},}\ }\BibitemShut {NoStop}%
\bibitem [{\citenamefont {Landau}\ and\ \citenamefont
  {Lifshitz}(2005)}]{Landau:2005fr}%
  \BibitemOpen
  \bibfield  {author} {\bibinfo {author} {\bibfnamefont {L.~D.}\ \bibnamefont
  {Landau}}\ and\ \bibinfo {author} {\bibfnamefont {E.~M.}\ \bibnamefont
  {Lifshitz}},\ }\href@noop {} {\emph {\bibinfo {title} {The Classical Theory
  of Fields}}},\ \bibinfo {edition} {fourth revised english}\ ed.,\ \bibinfo
  {series} {Course of Theoretical Physics}, Vol.~\bibinfo {volume} {2}\
  (\bibinfo  {publisher} {Butterworth-Heinemann},\ \bibinfo {year}
  {2005})\BibitemShut {NoStop}%
\bibitem [{\citenamefont {Max}(1973)}]{max:1480}%
  \BibitemOpen
  \bibfield  {author} {\bibinfo {author} {\bibfnamefont {C.~E.}\ \bibnamefont
  {Max}},\ }\href {\doibase 10.1063/1.1694545} {\bibfield  {journal} {\bibinfo
  {journal} {Physics of Fluids}\ }\textbf {\bibinfo {volume} {16}},\ \bibinfo
  {pages} {1480} (\bibinfo {year} {1973})}\BibitemShut {NoStop}%
\bibitem [{\citenamefont {Drake}\ \emph {et~al.}(1976)\citenamefont {Drake},
  \citenamefont {Lee},\ and\ \citenamefont {Tsintsadze}}]{Drake:1976fk}%
  \BibitemOpen
  \bibfield  {author} {\bibinfo {author} {\bibfnamefont {J.~F.}\ \bibnamefont
  {Drake}}, \bibinfo {author} {\bibfnamefont {Y.~C.}\ \bibnamefont {Lee}}, \
  and\ \bibinfo {author} {\bibfnamefont {N.~L.}\ \bibnamefont {Tsintsadze}},\
  }\href {\doibase 10.1103/PhysRevLett.36.31} {\bibfield  {journal} {\bibinfo
  {journal} {Phys. Rev. Lett.}\ }\textbf {\bibinfo {volume} {36}},\ \bibinfo
  {pages} {31} (\bibinfo {year} {1976})}\BibitemShut {NoStop}%
\bibitem [{\citenamefont {Gibbon}(2005)}]{Gibbon:2005ys}%
  \BibitemOpen
  \bibfield  {author} {\bibinfo {author} {\bibfnamefont {P.}~\bibnamefont
  {Gibbon}},\ }\href@noop {} {\emph {\bibinfo {title} {Short Pulse Laser
  Interactions with Matter: An Introduction}}}\ (\bibinfo  {publisher} {World
  Scientific Publication Company},\ \bibinfo {year} {2005})\BibitemShut
  {NoStop}%
\bibitem [{\citenamefont {Akhiezer}\ and\ \citenamefont
  {Polovin}(1956)}]{Akhiezer:1956mz}%
  \BibitemOpen
  \bibfield  {author} {\bibinfo {author} {\bibfnamefont {A.~I.}\ \bibnamefont
  {Akhiezer}}\ and\ \bibinfo {author} {\bibfnamefont {R.~V.}\ \bibnamefont
  {Polovin}},\ }\href@noop {} {\bibfield  {journal} {\bibinfo  {journal}
  {Soviet Physics, JETP}\ }\textbf {\bibinfo {volume} {3}},\ \bibinfo {pages}
  {696} (\bibinfo {year} {1956})}\BibitemShut {NoStop}%
\bibitem [{\citenamefont {Bulanov}\ \emph {et~al.}(2001)\citenamefont
  {Bulanov}, \citenamefont {Califano}, \citenamefont {Dudnikova}, \citenamefont
  {Esirkepov}, \citenamefont {Inovenkov}, \citenamefont {Kamenets},
  \citenamefont {Liseikina}, \citenamefont {Lontano}, \citenamefont {Mima},
  \citenamefont {Naumova}, \citenamefont {Nishihara}, \citenamefont {Pegoraro},
  \citenamefont {Ruhl}, \citenamefont {Sakharov}, \citenamefont {Sentoku},
  \citenamefont {Vshivkov},\ and\ \citenamefont
  {Zhakhovskii}}]{S.V.-Bulanov:2001kx}%
  \BibitemOpen
  \bibfield  {author} {\bibinfo {author} {\bibfnamefont {S.}~\bibnamefont
  {Bulanov}}, \bibinfo {author} {\bibfnamefont {F.}~\bibnamefont {Califano}},
  \bibinfo {author} {\bibfnamefont {G.}~\bibnamefont {Dudnikova}}, \bibinfo
  {author} {\bibfnamefont {T.}~\bibnamefont {Esirkepov}}, \bibinfo {author}
  {\bibfnamefont {I.}~\bibnamefont {Inovenkov}}, \bibinfo {author}
  {\bibfnamefont {F.}~\bibnamefont {Kamenets}}, \bibinfo {author}
  {\bibfnamefont {T.}~\bibnamefont {Liseikina}}, \bibinfo {author}
  {\bibfnamefont {M.}~\bibnamefont {Lontano}}, \bibinfo {author} {\bibfnamefont
  {K.}~\bibnamefont {Mima}}, \bibinfo {author} {\bibfnamefont {N.~M.}\
  \bibnamefont {Naumova}}, \bibinfo {author} {\bibfnamefont {K.}~\bibnamefont
  {Nishihara}}, \bibinfo {author} {\bibfnamefont {F.}~\bibnamefont {Pegoraro}},
  \bibinfo {author} {\bibfnamefont {H.}~\bibnamefont {Ruhl}}, \bibinfo {author}
  {\bibfnamefont {A.}~\bibnamefont {Sakharov}}, \bibinfo {author}
  {\bibfnamefont {Y.}~\bibnamefont {Sentoku}}, \bibinfo {author} {\bibfnamefont
  {V.}~\bibnamefont {Vshivkov}}, \ and\ \bibinfo {author} {\bibfnamefont
  {V.}~\bibnamefont {Zhakhovskii}},\ }\href@noop {} {\emph {\bibinfo {title}
  {Reviews of Plasma Physics}}},\ edited by\ \bibinfo {editor} {\bibfnamefont
  {V.~D.}\ \bibnamefont {Shafranov}},\ Vol.~\bibinfo {volume} {22}\ (\bibinfo
  {publisher} {Springer Berlin / Heidelberg},\ \bibinfo {year}
  {2001})\BibitemShut {NoStop}%
\bibitem [{Note1()}]{Note1}%
  \BibitemOpen
  \bibinfo {note} {One can also incorporate the radiation reaction term by
  appropriately modifying the plasma frequency, which essentially implies
  change in the laser pump wavevector arising due to the it's dispersion in the
  plasma.}\BibitemShut {Stop}%
\bibitem [{Note2()}]{Note2}%
  \BibitemOpen
  \bibinfo {note} {We justify retaining both modes in the dispersion relation
  later by calculating and comparing the frequency mismatch of the anti-Stokes
  mode with the growth rate of the FRS instability.}\BibitemShut {Stop}%
\end{thebibliography}

%

\end{document}